\documentstyle[aps,pra,multicol,epsfig]{revtex}

\def\ket#1{| #1\rangle}
\def\bra#1{\langle #1 |}

\newtheorem{assumption}{Assumption}

\title{
\rightline{ \small DAMTP-2002-5\normalsize } \rightline{}
\centerline{
Countering quantum noise with supplementary classical information}}
\author{Jonathan Barrett}
\address{Centre for Mathematical Sciences, Wilberforce Road, Cambridge
CB3 0WA}

\date{January 15, 2002}

\begin{document}
\maketitle

\begin{abstract}
We consider situations in which i) Alice wishes to send quantum
information to Bob via a noisy quantum channel, ii) Alice has a
classical description of the states she wishes to send and iii) Alice
can make use of a finite amount of noiseless classical
information. After setting up the problem in general, we focus
attention on one specific scenario in which Alice sends a known qubit
down a depolarizing channel along with a noiseless cbit. We describe a
protocol which we conjecture is optimal and calculate the average
fidelity obtained. A surprising amount of structure is revealed even
for this simple case which suggests that relationships between quantum
and classical information could in general be very intricate.
\vskip10pt
PACS number(s): 03.67.-a
\end{abstract}
\vskip10pt

\begin{multicols}{2}
\section{introduction}\label{intro}

In the study of quantum information theory it is often assumed that
classical information is effectively noiseless, free and unlimited. In this
context many problems become trivial. For example, consider a
situation in which Alice wants to `teleport' \cite{teleport} a quantum state, whose
identity is known to her, to Bob (this has come to be known as `remote
state preparation' \cite{rsp,pati,lo}). If classical information is
considered to be free, then no teleportation-type procedure is
actually needed. Alice can simply call Bob on the telephone and tell
him what the state is. If they don't care how long the call lasts then
Bob can construct a state arbitrarily close to Alice's original.

Remote state preparation becomes non-trivial if we wish to
restrict the amount of classical information that Alice can send to
Bob. Of course, if Alice and Bob share a perfect singlet (or one ebit), then they
can achieve perfect teleportation with the transmission of only two
classical bits (cbits). But in \cite{rsp}, it is shown that if Alice and
Bob don't mind using up a large amount of entanglement, then in the
asymptotic limit as the number of states being teleported tends to
infinity, they can get away with sending only one cbit per qubit and
still retain arbitrarily good fidelity. `Large amount' of entanglement
means that the amount needed increases exponentially with the number
of qubits being sent. It is also shown that this
exponential increase becomes a mere multiplying factor if we allow
cbits to be sent from Bob to Alice. Further, an upper bound is plotted
for how many cbits must be sent if we wish to use less than one ebit
per qubit transmitted. In \cite{devetak}, an optimal procedure is
given for this less-than-one-ebit case.

Other issues will arise if we consider all quantum channels to be
noisy and thus prevent the sharing of perfect singlets. We might
consider, in the context of some given situation, how the transmission
of extra cbits can offset this problem. Of course, if imperfect
singlets are shared then one option is always to try to distill better
singlets \cite{distillation1,distillation2}. But note that (i) in general it is not
possible to distill a perfect singlet from a finite number of mixed
states, so the resulting states are still noisy \cite{nowernerdistillation2}, (ii) some states only
admit distillation if collective operations are allowed, that is
operations on more than one pair at once, (for example this is true of
Werner states \cite{nowernerdistillation2,nowernerdistillation1})
and this may be impractical in a given situation and most importantly
(iii) distillation itself involves the sending of cbits and if this is
expensive, distillation may not be the best option.

It may rarely be the case that the sending of (relatively noiseless)
cbits is expensive compared with the sending of (potentially noisy)
qubits. Even if so, by assuming always that classical information is
effectively free and thereby not bothering to count it, we may miss out
on interesting theoretical relations between quantum and classical
information.

\section{The Problem}

With the above in mind, we consider the following problem. Alice and Bob are separated by a noisy quantum
channel. Alice sends into the channel some quantum state,
drawn from an ensemble $\{p_j,\ket{\psi_j}\}$ (where the state
$\ket{\psi_j}$ is drawn with probability $p_j$ and Alice and Bob both
know the ensemble). Alice is given a classical description of which
state went into the channel and can also send
$n$ noiseless classical bits which encode part of this information. Bob then performs some operation on the state which he receives
in an effort to undo the effects of the noise. If Bob's eventual
state, given that $\ket{\psi_j}$ was sent, is $\sigma_j$, then the
average fidelity is given by $\bar{F} = \sum_j p_j
\bra{\psi_j}\sigma_j \ket{\psi_j}$. We wish to describe a scheme for
Alice and Bob which will optimize this
quantity. 

We note that one might consider an alternative
scenario in which Alice is trying to prepare states remotely and can
generate any state she wants to be sent into the channel. The difference
lies in the fact that Alice may, in this alternative case, generate
and send a state which is different from the one which she wants
Bob ultimately to end up with. Here we do not investigate this
possibility but concentrate
only on the scenario in which the state entering
the channel is always identical with the state Alice wishes Bob to
prepare. This would be the case if Alice has no control over what goes
into the channel but is simply given a classical description. Or
if Alice sends the state she wants Bob to have into the channel
expecting it to be noiseless and only finds out about the noise later,
at which point she decides to send some additional classical information.

We start by describing how the most general possible scheme will
work. It is well known that the most general evolution which a quantum state
can undergo (assuming that if measurements are performed, their
results are to be averaged over)
corresponds to a completely positive trace-preserving map
\cite{book}. In the
Kraus representation, this can be written as:
\begin{equation}
\rho \longrightarrow \rho' = \sum_i M_i \rho M_i^{\dagger},
\end{equation}
where
\begin{equation}
\sum_i M_i^{\dagger} M_i = I
\end{equation}
and $I$ is the identity. We refer to such a map as a quantum
operation. Both the noise experienced by the state as it passes down
the channel and Bob's operation take this form. Now consider that the
experiment described above is repeated many times - each time, Alice
sends a quantum state down the noisy channel, and $n$ classical bits,
and Bob performs some quantum operation. Focus attention on those runs
of the experiment in which the classical bit-string has a certain
value, say $k$. We may as well regard Bob as performing the same
quantum operation on each of these runs. We use the fact that a
probabilistic mixture of quantum operations is itself a quantum
operation. So we can stipulate without loss of generality that which
quantum operation Bob applies depends deterministically on the values
of the $n$ cbits.

We can also stipulate without loss of generality that the values of
the $n$ cbits sent by Alice depend deterministically on which quantum state
she is sending. Suppose, to the contrary, that a particular quantum
state determines only probabilistically the values of the cbits. Then,
instead of regarding Alice and Bob as using a probabilistic scheme,
one might regard them as using one from several deterministic schemes,
with certain probabilities. But then the average fidelity obtained
will be the average over that obtained for each of the deterministic
schemes and we would do better simply to use whichever of these is the
best.

It follows from the above that we lose no generality if we restrict
ourselves to schemes which work as follows. The ensemble is divided up
into $2^n$ sub-ensembles. Alice uses the $n$ classical bits to tell
Bob which sub-ensemble the state she is sending lies in. Bob has a
choice of $2^n$ possible quantum operations to perform. Which one he
performs is determined by the values of the $n$ classical bits. The
problem is to find the scheme which leads to the maximum value for
$\bar{F}$.

We can
split this problem into two. The first part is to determine, for a
general ensemble of quantum states, $\{ p_i, \ket{\psi_i} \}$ which undergo some noise process of
the form $\rho \rightarrow {\cal S} (\rho )$, where ${\cal S}$ is a
quantum operation: What is the best
operation to perform in order to undo this noise as well as possible?
In other words, we wish to find an operation ${\cal T}$ such that
$\sum_i p_i \bra{\psi_i} {\cal T} ( {\cal S} ( \ket{\psi_i} \bra{\psi_i} )) \ket{\psi_i}$
is maximized. The second part is to determine the best way for Alice
to divide the initial ensemble into $2^n$ sub-ensembles, given that an
answer to the first part will determine for Bob an operation to
perform on each sub-ensemble.

Unfortunately, even the first of these appears to be a difficult
problem in itself. Some progress is made by Barnum and Knill in
\cite{barnumknill}, but they are concerned with maximizing
entanglement fidelity and their results are only valid for ensembles
of commuting density operators and so are not immediately useful for
our problem.

For the rest of this paper, we are less ambitious. We consider only a
very simple instance of the problem in which Alice sends just one cbit
and the pure states she sends are qubit states drawn from a
distribution which is uniform over the Bloch sphere. The noisy quantum
channel is a depolarizing channel, which acts as:
\begin{equation}
\rho \longrightarrow \alpha \rho + (1-\alpha ) \frac{I}{2},
\end{equation}
where $0\leq\alpha\leq1$. We will see that the solution even to this
seemingly trivial problem involves a surprising amount of structure -
suggesting that relationships between classical and quantum
information in general may well be very intricate.

\section{One qubit and one cbit}

Consider the scenario in which Alice sends to Bob a pure state drawn
from a uniform distribution over the Bloch sphere, which gets
depolarized on the way, and a single noiseless classical bit. From the
above, we know that Alice must divide the surface of the Bloch sphere
into two subsets, $S_0$ and $S_1$, which correspond to the cbit taking the value `0' or
`1'. We must then find, in each case, the optimal quantum operation
for Bob to perform, given that the depolarized qubit lies in that
particular subset.

  We begin by assuming that Alice divides up the Bloch sphere in the
following fashion:
\begin{assumption}\label{assumption}
For a general state, $\ket{\psi}$, we have that $\ket{\psi } \in S_0
\ {\rm iff} \ |\langle \psi |
0 \rangle |^2 \geq \cos^2 ( \beta / 2 )$, where $\ket{0}$ is some fixed basis
state corresponding to the point $(0,0,1)$, or the north pole, on the Bloch sphere and $0 \leq
\beta \leq \pi / 2$. Otherwise $\ket{\psi } \in S_1$.
\end{assumption}
We conjecture that this assumption leads to an optimal scheme (it
seems very likely, for example, that in the optimal scheme the sets
$S_0$ and $S_1$ will be simply connected and unlikely that the optimal
scheme will be less symmetric than the one presented). In the
rest of this section, we derive the optimal quantum operations for Bob
to perform, in the cases that $\ket{\psi } \in S_0$ and $\ket{\psi }
\in S_1$. 

It is helpful to write quantum operations in a different way. Suppose
that a general qubit density matrix is written
\begin{equation}
\rho = \frac12(I + \vec{r}.\vec{\sigma}),
\end{equation}
where $\vec{r}$ is a real 3-vector and $|\vec{r}| \leq 1$. Then, from
the fact that a quantum operation is linear, we can write it in the
form \cite{book}:
\begin{equation}
\vec{r} \longrightarrow \vec{r}' = A \vec{r} + \vec{b},
\end{equation}
where $A$ is a real $3 \times 3$ matrix and $\vec{b}$ is a real
3-vector. We have also automatically included the conditions that a
quantum operation must be trace-preserving and positive. The condition
of complete positivity imposes further constraints on $A$ and
$\vec{b}$. Of course we must also have that $|\vec{r}'| \leq 1 \
\forall \ \vec{r}'$. Further, we can write $A$ in the form $U.S$, where
$U$ is orthogonal (i.e. a rotation) and $S$ is symmetric. So we can
view a quantum operation as a deformation of the Bloch sphere along
principal axes determined by S, followed by a rotation, followed by a
translation.

Suppose now that $\ket{\psi} \in S_0$. Bob performs an operation
characterized by $A$ and $\vec{b}$. From the symmetry of the problem,
it follows that the fidelity obtained (averaged over all $\ket{\psi}$
such that $\ket{\psi} \in S_0$) is unchanged if Bob performs a
different operation, characterized by $A'$ and $\vec{b}'$, where
$A'=O(\theta ).A.O(\theta )^T$ and $\vec{b}'=O(\theta ).\vec{b}$ and
where $O(\theta )$ is a rotation of angle $\theta$ about the
z-axis. It follows from this that the fidelity is also unchanged if
Bob performs an operation characterized by $A''$ and $\vec{b}''$,
where:
\begin{equation}
A'' = \frac{1}{2\pi} \int_0^{2\pi} {\rm d} \theta \ O(\theta
).A.O(\theta )^T
\end{equation}
and
\begin{equation}
\vec{b}'' = \frac{1}{2\pi} \int_0^{2\pi} {\rm d} \theta \ O(\theta ).
\vec{b}.
\end{equation}
This means that without loss of generality, we can restrict Bob to
actions of the form $\vec{r} \rightarrow V.A. \vec{r} + \vec{b}$, where
$A = {\rm Diag}(\gamma, \gamma, \delta)$, $\vec{b} = (0,0,k)$ and $V$
is a fixed rotation about the z-axis. From
the condition that $|\vec{r}'| \leq 1 \ \forall \ \vec{r}'$, we get
$|\gamma|, |\delta|, |k| \leq 1$. Quantum operations are contractions
on the Bloch sphere.

Recall that the qubit which Bob receives has been depolarized. We can
write its density matrix in the form $\rho = 1/2 ( I + \alpha \, 
\vec{r}.\vec{\sigma} )$, where $|\vec{r}|=1$. Ideally, Bob would like
an operation which takes $\alpha \vec{r} \rightarrow \vec{r}$, at
least for those states belonging to $S_0$, but this
is not allowed (such an operation is not a contraction). Bob's
operation will in fact consist of a translation in the z-direction and
contractions parametrized by $\gamma$ and $\delta$. It is clear
geometrically that in the optimum scheme, $V=I$, where $I$ is the identity.

Our aim is now, for fixed $\alpha$, $\beta$ and $k$, to find the
optimum values of $\gamma$ and $\delta$, consistently with their
describing a genuine quantum operation (which, recall, must correspond
to a completely positive map on the set of density matrices). In fact,
one can show that complete positivity implies that:
\begin{eqnarray}
0&\leq&k\leq1, \\
0&\leq&\delta\leq1-k,
\end{eqnarray}
and
\begin{equation}
0\leq\gamma\leq\sqrt{1-k}.
\end{equation}
These conditions are necessary but not sufficient. The actual
derivation of these conditions is unenlightening, so we do not
reproduce it here.

It is easy to see now that Bob's best operation will be characterized
by setting $\gamma=\sqrt{1-k}$ and $\delta=1-k$. This gives:

\begin{equation}
A = {\rm Diag}(\sqrt{1-k},\sqrt{1-k},1-k)
\end{equation}
and
\begin{equation}
\vec{b} = (0,0,k).
\end{equation}
In fact, remarkably, this corresponds to an already well known quantum
operation, usually described as an `amplitude damping channel'. Amplitude damping is usually studied for its physical
relevance - it corresponds to many natural physical processes. For example, it may describe an atom coupled to a single mode of
electromagnetic radiation undergoing spontaneous emission, or a single
photon mode from which a photon may be scattered by a beam
splitter \cite{book}. This suggests that our scheme should be easily
implementable experimentally.

One can run through similar arguments for the case $\ket{\psi} \in
S_1$. Again, it turns out that Bob's optimal operation is essentially an amplitude
damping operation, except that in this case, the vector $\vec{b}$
will point in the opposite direction i.e. Bob's operation will involve
a translation of the Bloch sphere downwards, towards the south pole,
as well as some contraction. For the rest of this paper we calculate
the optimum fidelity that Bob can achieve for a given $\alpha$. For
fixed $\beta$, one can optimize over the value of $k$ separately for
the cases $\ket{\psi} \in S_0$ and $\ket{\psi} \in S_1$ (the optimum
value of $k$ may sometimes be zero implying that Bob's best operation
is to do nothing). One can then optimize over $\beta$.

\section{Achievable Fidelity}

After the action of the depolarizing channel and Bob's quantum
operation, we have that:
\begin{eqnarray}
\lefteqn{\vec{r}=\left(\begin{array}{c} \sin\theta\cos\phi \\
\sin\theta\sin\phi \\ \cos\theta \end{array} \right)}
\hspace{15pt} \nonumber \\
&& \longrightarrow \vec{r}'=\left( \begin{array}{c} \sqrt{1-k}\,\alpha
\,\sin\theta\cos\phi \\ \sqrt{1-k} \,\alpha \,\sin\theta\sin\phi \\
(1-k)\,\alpha \,\cos\theta + k \end{array}\right).
\end{eqnarray}
The average fidelity is given by:
\begin{eqnarray}
\lefteqn{\bar{F}=\frac{1}{4\pi}\int_0^{2\pi}\!{\rm
d}\phi\int_0^{\beta}\!{\rm d}\theta \ \sin{\theta} }
\hspace{15pt} \nonumber \\ && 
\frac12 \left(1+\alpha\sqrt{1-k}\sin^2\theta+\alpha(1-k)\cos^2\theta+k\cos\theta\right)
\nonumber \\ \lefteqn{+ \frac{1}{4\pi}\int_0^{2\pi}\!{\rm
d}\phi\int_{\beta}^{\pi}\!{\rm
d}\theta\ \sin\theta} \hspace{7pt} \nonumber \\ &&
\frac12 \left(1+\alpha\sqrt{1-k'}\sin^2\theta+\alpha(1-k')\cos^2\theta-k'\cos\theta\right),\nonumber
\\
\end{eqnarray}
where $k'$ is Bob's quantum operation parameter in the case that
$\ket{\psi} \in S_1$ and is defined so that $A =
{\rm Diag}(\sqrt{1-k'},\sqrt{1-k'},1-k')$ and $\vec{b}=(0,0,-k')$. 

Optimizing over $k$, $k'$ and $\beta$ numerically leads to the
graph shown in Figure \ref{fidelitygraph} which shows the achievable
fidelity for a depolarizing channel parametrized by $\alpha$. Also
shown on the graph is the fidelity obtained in the case that Alice
sends no cbit and Bob performs no quantum operation.

\begin{minipage}[t]{7.5cm}
\begin{figure}[htb]
\epsfig{file=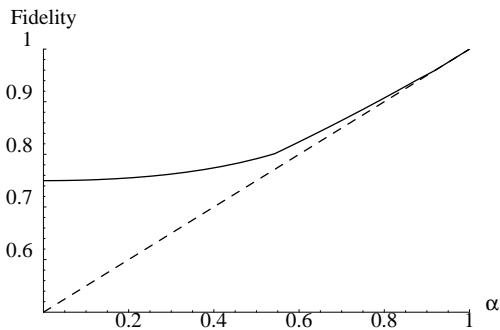}
\caption[Fidelity graph]{Average fidelity obtained for a depolarizing
channel parametrized by $\alpha$. Two cases are shown: (i) a cbit is
sent and Bob performs an `amplitude damping' operation (solid line) and (ii) Bob
performs no operation (dashed line).\label{fidelitygraph}}
\end{figure}
\end{minipage}
\vskip15pt
We finish this section by noting some features of the graph.
\begin{enumerate}
\item As we might expect, our scheme always yields an advantage when
compared with doing nothing.
\item If Alice can send to Bob one cbit but cannot use a quantum
channel, then the best obtainable fidelity is $3/4$ (Alice tells Bob
`upper' or `lower' hemisphere and Bob prepares a state which is spin
up or spin down accordingly). With our scheme we have $\bar{F} > 3/4$
if $\alpha > 0$. Thus the quantum channel is some use for any $\alpha
> 0$.
\item There is a kink in the graph at $\alpha \approx 0.54$. Further
numerical investigations reveal why this is the case. Denote the
optimum value of $\beta$ (the angle which describes how Alice is
dividing up the Bloch sphere) by $\beta_{opt}$. Below this value
of $\alpha$, we have $\beta_{opt}=\pi/2$. At $\alpha\approx0.54$,
$\beta_{opt}$ suddenly jumps to $\sim1.1$ and then decreases as
$\alpha$ increases. This is shown in Figure \ref{anglegraph}.

\begin{minipage}[t]{7.5cm}
\begin{figure}[htb]
\epsfig{file=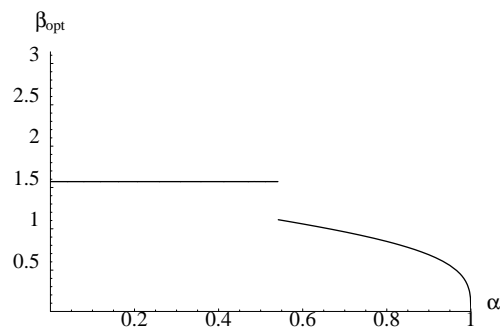}
\caption[Angle graph]{Optimal value of $\beta$ plotted against
$\alpha$, where $\beta$ characterizes how Alice divides up the Bloch sphere
(see assumption \ref{assumption}). Note that the graph is disjoint at
$\alpha \approx 0.54$.\label{anglegraph}}
\end{figure}
\end{minipage}
\vskip15pt
\item In the $\alpha < \ \sim\!0.54$ region, where $\beta_{opt}=\pi/2$, we
can calculate $\bar{F}$ analytically yielding
$\bar{F}=\frac34+\frac13\frac{\alpha^2}{3-2\alpha}$.
\item As $\alpha\rightarrow0$, Bob's operation tends towards a simple
`swap' operation which maps all points in the Bloch sphere to one of
the poles depending on which hemisphere the qubit lies in.
\item If $\alpha < \ \sim\!0.72$, then $\bar{F}< \ \sim\!0.872$ and Alice and
Bob, if they can, would do better to use a protocol due to Gisin in which Alice
sends two noiseless cbits and no quantum information \cite{gisin}.

\end{enumerate}

\section{Conclusion}

We have considered situations in which Alice and Bob wish to use
noiseless classical information to offset quantum noise - a kind of
error correction. An important feature is that Alice possesses a
classical description of the quantum states she wishes to send. After
considering these situations in generality, we turned to consider a
very specific scenario in which Alice sends one qubit which passes
through a depolarizing channel accompanied by a noiseless classical
bit. We described a scheme which we conjecture is optimal which
involves Alice dividing up the Bloch sphere as in assumption
\ref{assumption} above and Bob performing `amplitude damping'
operations. Our results for this scheme were obtained by brute
force. Clearly a more principled approach is desirable. One idea might
be to regard the depolarization as actually coming about through the
actions of an eavesdropper, Eve. Eve gains some information about the
quantum state passing through and must therefore gain some information
about its identity. It follows that even after Bob's recovery
operation, some disturbance to the state is inevitable
\cite{fuchs}. This way, one might be able to derive an upper bound on
Bob's achievable fidelity for more general scenarios than the one
considered here.

\leftline{\bf Acknowledgments}

I am grateful to Trinity College, Cambridge for support, CERN for
hospitality and to the European grant EQUIP for partial support.
I would like to thank Adrian Kent and Sandu Popescu for useful discussions.

\end{multicols}

\begin{thebibliography}{99}

\bibitem{teleport} 
C.~H.~Bennett, G.~Brassard, C.~Crepeau, {\it et al.},
Phys. Rev. Lett. {\bf 70}, 1895 (1993).
\bibitem{rsp}
C.~H.~Bennett, D.~P.~DiVincenzo, P.~W.~Shor, {\it et al.},
Phys. Rev. Lett. {\bf 87}, 077902 (2001).
\bibitem{pati}
A.~K.~Pati, Phys. Rev. A {\bf 63}, 014302 (2001). 
\bibitem{lo}
H-K.~Lo, Phys. Rev. A {\bf 62}, 012313 (2000).
\bibitem{devetak}
I.~Devetak and T.~Berger, Phys. Rev. Lett. {\bf 87}, 197901 (2001).
\bibitem{distillation1}
C.~H.~Bennett, H.~J.~Bernstein, S.~Popescu, {\it et al.}, Phys. Rev. A
{\bf 53}, 2046 (1996).
\bibitem{distillation2}
C.~H.~Bennett, G.~Brassard, S.~Popescu, {\it et al.},
Phys. Rev. Lett. {\bf 76}, 722 (1996).
\bibitem{nowernerdistillation2}
A.~Kent, Phys. Rev. Lett. {\bf 81}, 2839 (1998).
\bibitem{nowernerdistillation1}
N.~Linden, S.~Massar and S.~Popescu, Phys. Rev. Lett. {\bf 81}, 3279 (1998).
\bibitem{book}
M.~Nielsen and I.~Chuang, ``Quantum Computation and Quantum
Information'' (Cambridge University Press, 2000).
\bibitem{barnumknill}
H.~Barnum and E.~Knill, quant-ph/0004088.
\bibitem{gisin}
N.~Gisin, Phys. Lett. A {\bf 210}, 157 (1996).
\bibitem{fuchs}
C.~Fuchs, Fortschr. Phys. {\bf 46}, 535 (1998).

\end{thebibliography}
\end{document}